\documentstyle[12pt,epsfig]{article}
\newcommand{\be}{\begin{equation}}
\newcommand{\ee}{\end{equation}}
\newcommand{\bea}{\begin{eqnarray}}
\newcommand{\eea}{\end{eqnarray}}
\newcommand{\bdm}{\begin{displaymath}}
\newcommand{\edm}{\end{displaymath}}

\begin{document}
\begin{titlepage}
\rule{0em}{2em}
\begin{center}
{\LARGE {\bf $0^{++}$-Glueball/$q \bar q$-State Mixing in the
Mass Region near 1500 MeV}} \\
 \vspace{1.5cm}
 A.V. Anisovich, V.V.Anisovich and A.V.Sarantsev\\
 St.Petersburg Nuclear Physics Institute\\
 Gatchina, St. Petersburg 188350 \\Russia\\
\vspace{1cm}
{\bf Abstract} \\
 \vspace{2em}

\parbox{30em}{Basing on the results of the K-matrix fit of
$(IJ^{PC}=00^{++})$ wave \cite{Kmatrix}, we analyze  analytic
structure of the amplitude and $q \bar q$/glueball content of
 resonances in the mass region 1200-1900 MeV, where an extra
state for  $q \bar q$-systematics exists being a good candidate
for the lightest scalar glueball. Our analysis shows that the
pure glueball state  dispersed over three  resonances:
$f_0(1300)$, $f_0(1500)$ and $f_0(1530^{+90}_{-250})$, while
the glueball
admixture in $f_0(1750)$ is small. The broad resonance
$f_0(1530^{+90}_{-250})$ is the descendant of the lightest pure
glueball.  The mass of pure glueball is $1630\pm {70\atop 30}$
MeV, in agreement with  Lattice calculation results
\cite{Close-Bali, Sexton}.}

\end{center}
\end{titlepage}

In ref. \cite{Kmatrix} the K-matrix analysis of $00^{++}$-wave
has been performed in the mass region 500-1900 MeV for the
channels $\pi\pi,\; K \bar K,\; \eta\eta,\; \eta\eta'$ and
$\pi\pi\pi\pi$. Simultaneous fit of the data of
refs. [4-8] fixes five K-matrix
poles (or five bare states, in the terminology of
ref. \cite{Kmatrix}).  Only two of them are definitely $s \bar
s$-rich states:  $f_0^{bare}(720 \pm 100)$ and $f_0^{bare}(1830
\pm 30)$. For other three states, $f_0^{bare}(1230 \pm 50)$,
$f_0^{bare}(1260 \pm 30)$ and $f_0^{bare}(1600 \pm 50)$: two of
them are natural $q \bar q$-nonet partners for $f_0^{bare}(720)$
and $f_0^{bare}(1830)$, while one state is extra for $q \bar
q$-systematics. The $q \bar q$/glueball content of meson states
reveals itself in coupling ratios for decays into channels
$\pi\pi,\; K \bar K,\; \eta\eta$ and $\eta\eta'$
\cite{Gershtein, Amsler, Anis95}. By use of these coupling
ratios, the analysis of ref. \cite{Kmatrix} gives two solutions
which describe the data set well:\\
Solution I:
\begin{description}
\item[~~~] $f_0^{bare}(720)$ and $f_0^{bare}(1260)$ are $1^3P_0$ nonet
partners,
\item[~~~] $f_0^{bare}(1600)$ and $f_0^{bare}(1810)$ are $2^3P_0$
nonet partners,
\item[~~~] $f_0^{bare}(1230)$ is a glueball;
\end{description}
Solution II:
\begin{description}
\item[~~~] $f_0^{bare}(720)$ and $f_0^{bare}(1260)$ are $1^3P_0$ nonet
partners,
\item[~~~] $f_0^{bare}(1230)$ and $f_0^{bare}(1810)$ are $2^3P_0$
nonet partners,
\item[~~~] $f_0^{bare}(1600)$ is a glueball.
\end{description}

Physical states are mixtures of bare states which
occur in the K-matrix formalism via transitions of bare states
into meson channels (in the analysis of ref. \cite{Kmatrix}:
$\pi\pi,\; K \bar K,\; \eta\eta,\; \eta\eta'$ and
$4\pi$). In the mass region 1200-1650 MeV which is key region
for determination of the lightest glueball, the
$00^{++}$-amplitude has four poles at the following complex
masses (in MeV)  \cite{Kmatrix}:\\
\be
\begin{array}{lclcl}
(1300 \pm 20)& -& i (120 \pm 20)& \rightarrow &f_0(1300)\\
(1499 \pm 8)& - &i (65 \pm 10)& \rightarrow &f_0(1500)\\
(1530^{+90}_{-250})& -& i (560 \pm 140) &\rightarrow &
f_0(1530^{+90}_{-250})\\
(1780 \pm 30)& -& i (125 \pm 70) &\rightarrow &f_0(1780)
\end{array}
\label{1}
\ee
Each of these states is a mixture of $q \bar q$ and glueball
components. In order to reconstruct $q \bar q$/glueball content
of the resonances, we have performed here a re-analysis of the
$00^{++}$-amplitude in the region 1200-1650 MeV using the
language of  $q \bar q$ and glueball states.

\section{Glueball propagator}

Mixing of $q \bar q$-states with glueball is due to the
processes shown in fig. 1a,b: gluons of the glueball produce a
$q \bar q$-pair,  fig. 1a; the produced quarks interact by gluon
exchanges, fig. 1b. According to the rules of the 1/N-expansion
\cite{t'Hooft}, the main contribution into the interaction block
is given by planar diagrams. Saturating the $q \bar q$-scattering
 block of fig. 1b by $q \bar q$ states, we represent the
diagrams of fig. 1b as a set of diagrams of the type shown in
figs. 1d and 1e. The sum of diagrams of fig. 1c, 1d, 1e, and so
on gives the glueball propagator with $q \bar q$-state mixing
taken into account. The mixtures of the pure glueball state and
input $q \bar q$-states are determined by the quark loop
transition diagrams, $B_{ab}(s)$, ($s=p^2$ is glueball
four-momentum squared) which have the following form in the
light cone variables:
\be
B_{ab}(s)=\frac{1}{(2\pi)^3}
\int\limits_{0}^{1} \frac{dx}{x} \int d^2 k_{\bot}\;
\frac{g_a(s') g_b(s')}{s'-s-i0}\; 2(s'-4m^2).
\label{2}
\ee
Here $s'=\frac{m^2+k^2_{\bot}}{x(1-x)}$, $g_a$ and $g_b$ are the
vertices of the transitions $state\;a \rightarrow q \bar q$
and $state\;b \rightarrow q \bar q$, and $m$ is  quark
mass. Factor $2(s'-4m^2)$ is determined by the spin
structure of the quark loop diagram:
$Tr[(\hat k +m)(-\hat p + \hat k +m)]= 2(s'-4m^2)$.

To analyse analytic structure of the
$00^{++}$-amplitude in the mass region of the resonances of
eq. (\ref{1}), let us introduce a $4\times4$
propagator matrix, $D_{ab}(s)$, which describes the transition
$state \; a \rightarrow state\;b$ with $a,b=1,2,3,4$, in accordance
with the number of  investigated states. The
diagrams of  fig. 1 type give:
\be
D^{-1}_{ab}(s)=(m_a^2-s)\delta_{ab}-B_{ab}(s)
\label{3}
\ee
Here $m_a$ is an input mass for the state $a$; in the case of
the glueball it is the mass of  pure gluonic glueball,
$\delta_{ab}$ is  unit matrix, and $B_{ab}(s)$ is given
by eq. (2). The zeros of the determinant
\be
\Pi(s)=det|(m_a^2-s)\delta_{ab}-B_{ab}(s)|
\label{x}
\ee
determine the complex masses of physical states:
in the case under investigation, they are given by
eq. (\ref{1}). Let us denote them as $M_A$, $M_B$ $M_C$ and $M_D$.
Then, in the vicinity of  $s=M_A^2$, $D_{ab}(s)$ is described
by the pole term only:
\be
D_{ab}(s \sim M_A^2) \simeq N_A\; \frac{\alpha_a
\alpha_b}{M^2_A-s},
\label{5}
\ee
where four coefficients $\alpha_1, \alpha_2, \alpha_3,
\alpha_4$ satisfy the constraint
\be
\alpha^2_1 + \alpha^2_2 +\alpha^2_3
+\alpha^2_4=1
\ee
 and determine the probabilities of the input states (1,2,3,4)
in the physical state $A$; $N_A$ is a normalization factor
common for all $D_{ab}$.

In order to take into account the flavor content in the quark loop
diagrams which is omitted in eq. (\ref{2}), the following
replacement should be done:
\be
B_{ab}(s) \rightarrow  cos\phi_a cos\phi_b B_{ab}^{(n \bar
n)}(s) +
sin\phi_a sin\phi_b B_{ab}^{(s \bar s)}(s),
\label{6}
\ee
where $|a \rangle =cos \phi_a\, n \bar n + sin \phi_a\, s \bar
s$ and
$|b \rangle =cos \phi_b\, n \bar n + sin \phi_b\, s \bar s$,
while $B_{ab}^{(n \bar n)}$ and $B_{ab}^{(s \bar s)}$ refer
to  loop diagrams with non-strange and strange quarks,
$n\bar{n}=(u\bar{u}+d\bar{d})/\sqrt{2}$  and
 $s\bar{s}$.  For  pure glueball state, $a=glueball$, the
effective mixing angles are
determined by relative probabilities of the production
of non-strange and strange quarks by gluons, $u \bar u\;:\;d \bar d \;:\;s
\bar s= 1\;:\;1 \;:\;\lambda$, so that
$tg\,\phi_{glueball}=\sqrt{\lambda /2}$.
Experimental data give $\lambda \simeq 0.5$ \cite{Anis95,Anis90}: this value
corresponds to $\phi_{glueball} \simeq 25^\circ$. Mixing angles for $1^3P_0
q\bar{q}$ and $2^3P_0 q \bar q$ states were found in ref. \cite{Kmatrix}.

\section {Fit of the $00^{++}$ amplitude}

For the calculation of  loop diagrams, eq. (\ref{2}), we
should fix the vertices $g_a(s)$. We parametrize the vertices
for the transition $state \;a \rightarrow n \bar n$
in a simple form:\\
\be
\begin{array}{ll}
1^3P_0~q \bar q-state:&
g_1(s)= \gamma_1\; \sqrt[4] s \;\frac{k_1^2+\sigma_1}{k^2+\sigma_1};\\
~&\\
2^3P_0~q \bar q-first\; state:&
g_2(s)= \gamma_2\; \sqrt[4] s \;\left [
\frac{k_2^2+\sigma_2}{k^2+\sigma_2} - d
\frac{k_2^2+\sigma_2}{k^2+\sigma_2+h} \right ];\\
~&\\
Glueball:&
g_3(s)= \gamma_3\; \sqrt[4] s \;\frac{k_3^2+\sigma_3}{k^2+\sigma_3};\\
~&\\
2^3P_0~q \bar q-second\; state:&
g_4(s)=g_2(s).
\end{array}
\label{7}
\ee
Here $k^2=\frac{s}{4} -m^2$ and $k_a^2=\frac{m_a^2}{4} -m^2$; $m_a$,
$\gamma _a$ and $\sigma_a$ are parameters. Factor $d$ is due to
 orthogonality of the $1^3P_0 q\bar q$ and $2^3P_0 q\bar q$
states: we put $Re\;B_{12}(s_0)=0$ at $\sqrt{s_0}=1.5$ GeV. (In the case
of
$s$-dependent $B$-functions the orthogonality requirement for loop
transition diagrams cannot be fixed at all values of $s$).

The parameters $m_a$, $\sigma_a$, $h$ and $\gamma_a$ ($a=1,2,3$)
are to be determined by  masses and widths of the physical
resonances of eq. (\ref{1}).
However, the masses $m_a$
are approximately fixed by  the K-matrix fit of
ref. \cite{Kmatrix}, where masses of the K-matrix poles,
$M_a^{bare}$, are determined: $(M_a^{bare})^2 \simeq m_a^2-
B_{aa}^2((M_a^{bare})^2)$. Let us stress that
$m_3$ is the mass of  pure gluonic glueball which
is a subject of Lattice QCD calculation.

Parameters which are found in our fit of the $00^{++}$ amplitude in
the mass region 1200-1900 MeV are given in Table 1.
Using these parameters, we calculate the couplings $\alpha_a$
which are introduced by eqs. (5) and (6): these couplings determine
 relative weight of the initial state $a$ in the physical resonance
$A$:
\be
W_a(A)= |\alpha_a|^2
\label{10}
\ee
The probabilities  $W_a$ are given in Table 2 together
with masses of  physical resonances, $M_A$, and  masses of
 input states, $m_a$.

\section{Glueball/$q \bar q$-state mixing}

In order to analyze the dynamics of the glueball/$q \bar q$
mixing, we use the following method: in the final formulae the
vertices are replaced in a way:
\be
g_a(s) \rightarrow  \xi g_a(s),
\ee
with a factor $\xi$ running in the interval $0 \leq \xi \leq 1$.
The case $\xi=0$ corresponds to switching off mixing of the input
states. The input states are stable in this case, and
corresponding poles of the amplitude are at $s_a=m_a^2$. Fig. 2
shows the pole position at $\xi=0$ for  solution I (fig. 2a)
and  solution II (fig. 2b). For  glueball state $m_3$ is
the mass of a pure glueball, without $q \bar q$ degrees of
freedom. In  solution I the pure-glueball mass is equal to
\be
m_{pure\;glueball}(Solution\;I)=1225\;MeV,
\ee
that definitely disagrees with the Lattice-Gluodynamics
calculations for the lightest glueball. In solution II
\be
m_{pure\;glueball}(Solution\;II)=1633\;MeV.
\ee
This value is in a good agreement with recent
Lattice-Gluodynamics results:\\
1570$\pm$85(stat)$\pm$100(syst) MeV \cite{Close-Bali} and
1707$\pm$64 MeV \cite{Sexton}.
With increasing $\xi$ the poles are shifted into  lower part
of the complex mass plane. Let us discuss in detail the
solution II which is consistent with Lattice result.

At $\xi \simeq 0.1-0.5$ the glueball state of     solution II is
mixing mainly with $2^3P_0$ $q \bar q$-state, at $\xi \simeq
0.8-1.0$ the mixture with $1^3P_0$ $q \bar q$-state  becomes
significant. As a result, the descendant of the pure glueball
state has the mass $M=1450-i450$ MeV. Its gluonic content is
47\% (see Table 1). We should emphasize: the
definition of $W_a$  suggests that
$\sum_{A=1,2,3,4} W_{glueball}(A) \neq 1$ because of the
s-dependent $B_{ab}$
in the propagator matrix.

Hypothesis that the lightest scalar glueball is strongly mixed
with neighbouring $q\bar q$ states  was
discussed previously (see refs.
 \cite{AC-PR}, \cite{Uspekhi}, and
 references therein).
However, the attempts to reproduce a quantitative picture
of the glueball/$q\bar q$-state mixing and the mass shifts
caused by this mixing could not be
successful within standard quantum mechanics approach
that misses two phenomena:
\\ (i) Glueball/$q\bar q$-state mixing
described by propagator matrix can give both a repulsion of the mixed
levels, as in the standard quantum mechanics, and an attraction of them.
The latter effect may happen
because the loop diagrams $B_{ab}$ are complex
magnitudes, and the imaginary parts $ImB_{ab}$ are rather large
in the region 1500
MeV.
\\ (ii) Overlapping resonances yield a repulsion of the amplitude
pole positions along imaginary-s axis.  In the case of
full overlap of two resonances the width of one
state tends to zero, while the width of the second state
tends to the sum of
the widths of initial states, $\Gamma_{first} \simeq 0$ and $\Gamma_{second}
\simeq \Gamma_1+ \Gamma_2$. For three overlapping resonances the widths of two
states tend to zero,
 while the width of the third state accumulates the widths
of all initial resonances, $\Gamma_{third} \simeq \Gamma_1+ \Gamma_2 +
\Gamma_3$.

Therefore, in the case of  nearly overlapping
resonances, what occurs  in the region near 1500 MeV, it
is inevitable to have one resonance with a large width. It is
also natural that it is the glueball descendant with large width:
the glueball mixes with the neighbouring $1^3P_0$ $q \bar q$ and
$2^3P_0$ $q \bar q$ states, which are both $n \bar n$ rich,
without suppression.

\section{Problems}

Despite the fact that the lightest scalar glueball is now
understandable  in its principle points, there are problems
which need to be clarified. First, it is necessary to
estimate the influence of the more distant resonance,
$f_0(980)$. Second, the analysis should be
repeated in terms of hadron states, without using the language
of $q \bar q$-states. Such an analysis would give the
possibility to check the idea of quark---hadron duality in the
mass region 1000-2000 MeV which is used here.

There is one more problem: K-matrix analysis of $00^{++}$ wave
\cite{Kmatrix} provides two solutions. Correspondingly, analyzing
here two variants we
rejected one solution basing on the results of Lattice-Gluodynamics
calculations. The problem is if it is possible to discriminate between
solution I and solution II and what type of experimental data
are needed for that.

\section{Conclusion}

The lightest gluodynamic glueball is dispersed over  neighbouring
resonances mixing mainly to $1^3P_0$ $q \bar q$ and
$2^3P_0$ $q \bar q$ states. With this mixing the glueball
descendant transforms into  broad resonance,
$f_0(1530^{+90}_{-250})$. This resonance contains (40-50)$\%$
of the glueball component. Another part of the glueball
component is shared between comparatively narrow resonances,
$f_0(1300)$ and $f_0(1500)$ which are descendants of $1^3P_0$
$q \bar q$ and $2^3P_0$ $q \bar q$ states.

We thank T. Barnes, D.V. Bugg, F.E. Close, L.G. Dakhno, L. Montanet
and Yu.D. Prokoshkin for useful discussions. This work was
supported by RFFI grant N96-02-17934 and INTAS-RFBR grant N95-0267.

\newpage
\begin{center}
Table 1\\
\vskip 0.3cm
Masses of the initial states,  coupling constants
and $q\bar q$/glueball content of  physical states.
\vskip 0.3cm
\begin{tabular}{|c|c|c|c|c|}
\hline
\multicolumn{5}{|c|}{Solution I}  \\
\hline
~ & $1^3P_0$ & $2^3P_0$ & ~ & $2^3P_0$ \\
Initial state & $n\bar n$-rich & $n\bar n$-rich &
Glueball & $s\bar s$-rich\\
~ &$\phi_1=18^\circ$ &$\phi_2=-6^\circ$  & $\phi_3=25^\circ$ &
$\phi_4=84^\circ$\\
\hline
$m_i$ (GeV) & 1.457 & 1.536 & 1.230 & 1.750 \\
$\gamma_i$ (GeV$^{3/4}$) & 0.708 & 1.471 & 0.453 & 1.471 \\
$\sigma_i$ (GeV$^2$) & 0.075 & 0.225 & 0.375 & 0.225 \\
\hline
$W[f_0(1300)]$ & 32\% & 12\%  & 55\% & 1\%\\
$1.300-i0.115$ (GeV) & ~ &~ & ~& \\
$W[f_0(1500)]$ & 25\% & 70\%  & 3\% & 2\%\\
$1.500-i0.065$ (GeV) & ~ &~ & ~& \\
$W[f_0(1530)]$ & 44\% & 24\%  & 27\% & 4\%\\
$1.450-i0.450$ (GeV) & ~ &~ & ~& \\
$W[f_0(1780)]$ & 1\% &  1\%   & --    & 98\%\\
$1.780-i0.085$ (GeV) & ~ &~ & ~& \\
\hline
\multicolumn{5}{|c|}{$h=0.25$ GeV$^2$,~~~$d=1.01$}  \\
\hline
\hline
\multicolumn{5}{|c|}{Solution II}  \\
\hline
~ & $1^3P_0$ & $2^3P_0$ & ~ & $2^3P_0$ \\
Initial state & $n\bar n$-rich & $n\bar n$-rich &
Glueball & $s\bar s$-rich\\
~ &$\phi_1=18^\circ$ &$\phi_2=35^\circ$  & $\phi_3=25^\circ$ &
$\phi_4=-55^\circ$\\
\hline
$m_i$ (GeV)& 1.107 & 1.566 & 1.633 & 1.702 \\
$\gamma_i$ (GeV$^{3/4}$)& 0.512 & 0.994 & 0.446 & 0.994 \\
$\sigma_i$ (GeV$^2$) & 0.175 & 0.275 & 0.375 & 0.275 \\
\hline
$W[f_0(1300)]$ & 35\% & 26\%  & 38\% & 0.4\%\\
$1.300-i0.115$ (GeV) & ~ &~ & ~& \\
$W[f_0(1500)]$ & 1\% & 64\%  & 35\%  & 0.4\%\\
$1.500-i0.065$ (GeV) & ~ &~ & ~& \\
$W[f_0(1530)]$ & 12\% & 41\%  & 47\% & 0.3\%\\
$1.450-i0.450$ (GeV) & ~ &~ & ~& \\
$W[f_0(1780)]$ & 0.1\%  &  0.2\%   & 0.2\%   & 99.5\%\\
$1.750-i0.100$ (GeV) & ~ &~ & ~& \\
\hline
\multicolumn{5}{|c|}{$h=0.625$ GeV$^2$,~~~$d=1.16$}  \\
\hline
\end{tabular}
\end{center}

\newpage
\begin{center}
\epsfig{file=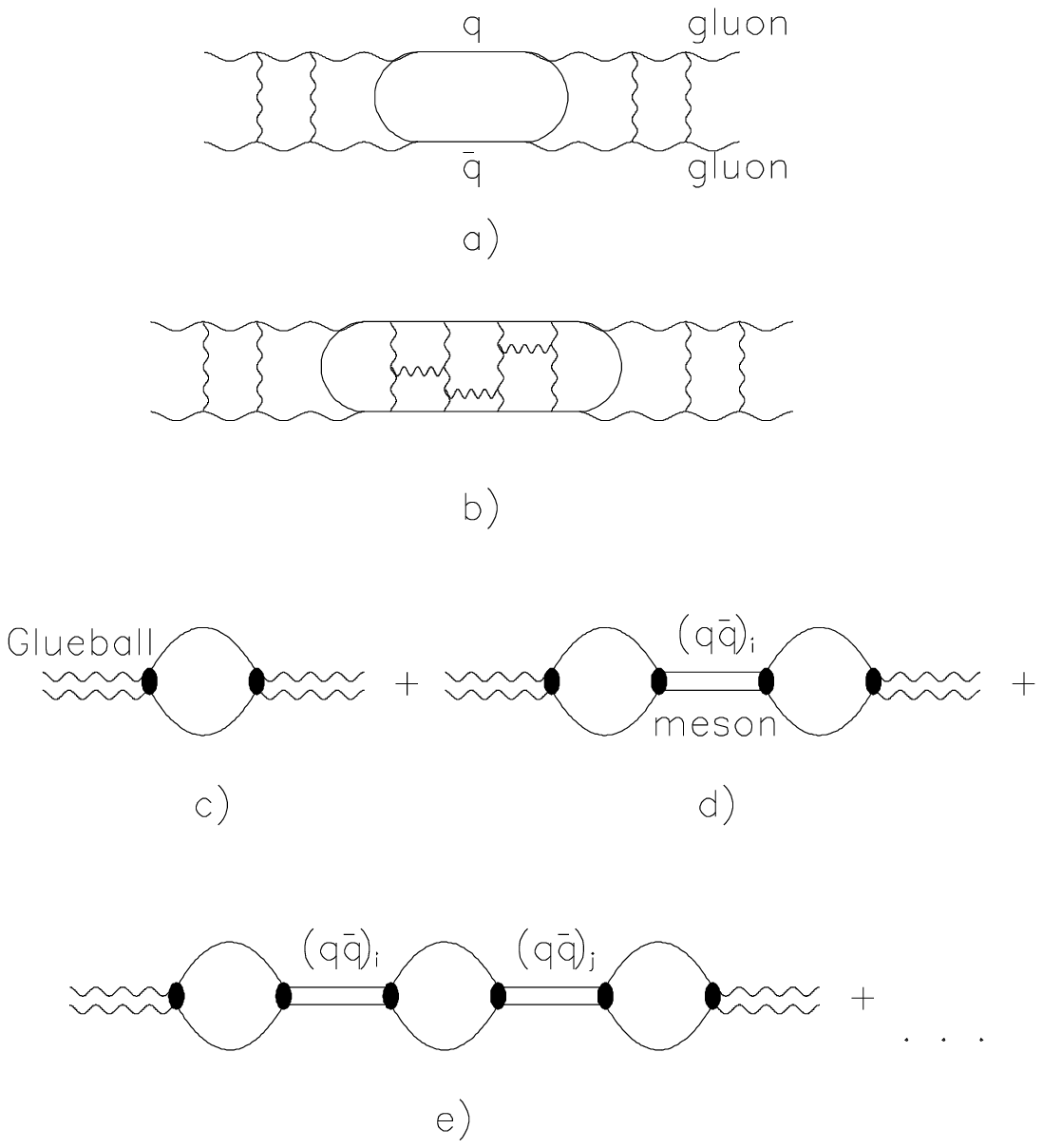,width=14cm}\\
Fig. 1. Diagrams which provide the glueball/$q\bar q$ mixing.
\end{center}

\newpage
\begin{center}
\epsfig{file=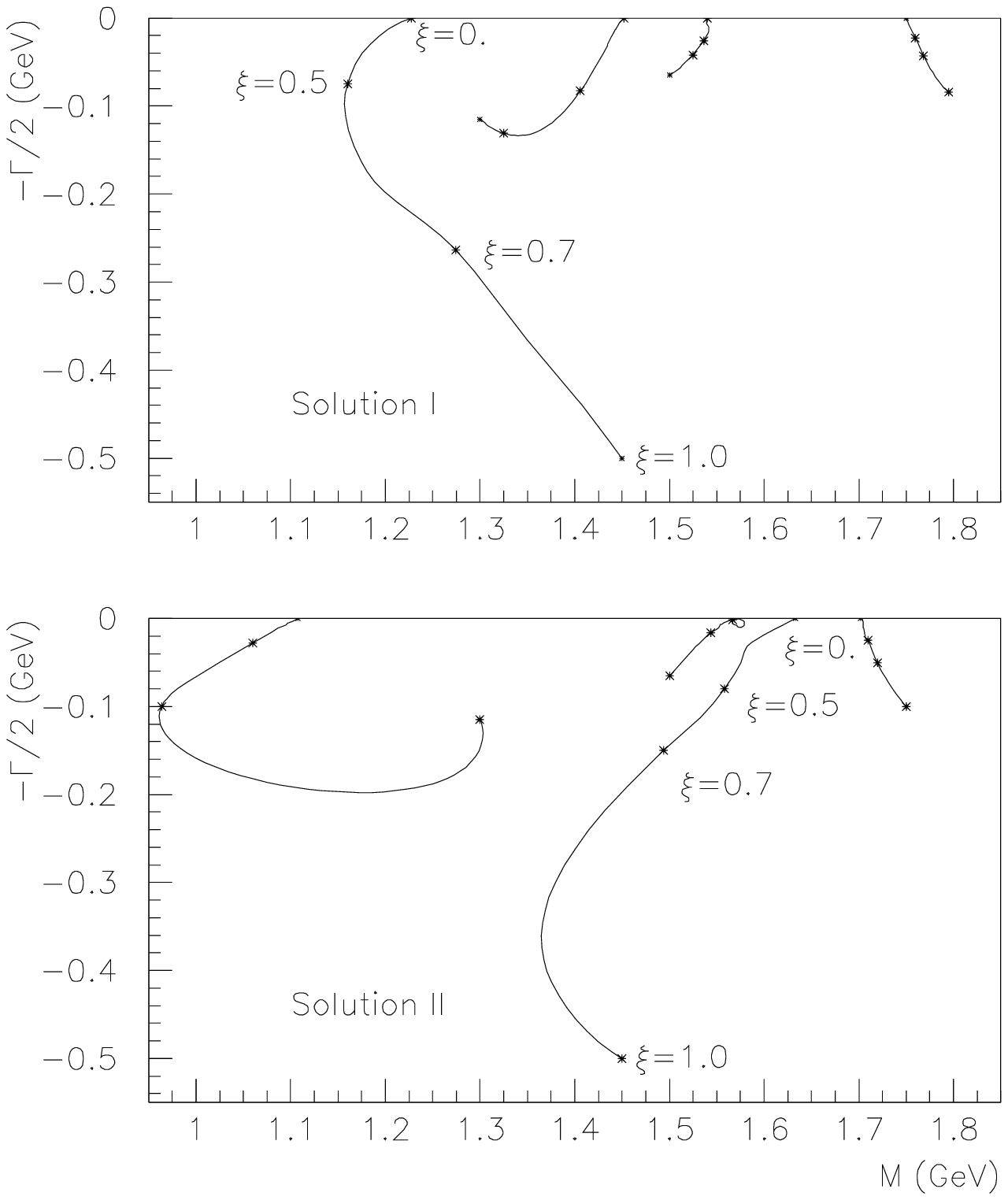,width=14cm}\\
Fig. 2. Complex-$\sqrt{s}$ plane
$(M=Re\sqrt{s},\;\;-\Gamma/2=Im\sqrt{s})$:  location of $00^{++}$
amplitude poles after replacing $g_a \to \xi g_a$.  The case $\xi=0$
gives the positions of masses of input $q\bar q$ states and gluodynamic
glueball; $\xi=1$ corresponds to the real case.  \end{center}

\end{document}